# WHY NOT A SUPERFLUID SOLID?

P W Anderson, Princeton University

ABSTRACT

The question of whether a Bose solid can have a superfluid fraction in the absence of interstitials, holes or other defects is discussed. A possible scenario which may accommodate this possibility is proposed, based on a Hartree-Fock treatment of the quantum solid.

___________________________________  ____________

Moses Chan and coworkers[1] have recently found experimental evidence that below about 200 millidegrees K solid helium has a non-rotating (superfluid) component amounting to roughly 1% of its moment of inertia. Many experimental checks failed to find any evidence of defects or grain boundary phenomena in the samples; solid He normally grows in very defect-free crystals  In addition, it seems to us that the fact that He in porous substrates does not behave appreciably differently is strong evidence against any defect explanation.

Theories of quantum solids come in many varieties.  Much of their complication comes from the attempt to take into account the large zero-point amplitude of the phonons, which carries them outside of the linear range of the interatomic potential and makes numerical computation very difficult.  Here however we are not very interested in numerical accuracy, and instead want to know whether superfluid behavior is possible in principle.  For these purposes the simple Hartree-Fock theory sketched in my book[2] should suffice, in that it shows how a Bose solid could occur and gives an explicit wave function for it.

The wave function proposed in ref 2 is

$$\Psi = (\prod_i c_i^*)\Psi_{vac} \quad [1]$$

where the c's refer to a set of localized self-consistent boson operators referring to orbitals localized at the sites i of a lattice:

$$c_i^* = \int d^3 r \phi_i(r)\psi^*(r) \quad [2]$$

$\psi^*(r)$ is the boson creation operator. I showed in reference 2 that the wave functions $\phi_i(r)$ satisfy a self-consistency equation obtained as the Hartree-Fock equation for hole excitations (which in the Bose case is not the same as that for particle excitations.) If the potential is sufficiently repulsive, this equation can have a self-consistent localized potential well because, in the boson case, for holes the exchange term adds to the self-consistent potential rather than compensating it as for fermions. For particles, on the other hand, the effective potential is perfectly periodic and cannot have a bound state—hence there is an energy gap separating the particle (interstitial) states from the hole (vacancy) ones, and the solid is stable at a density fixed (hypothetically) by optimizing the energy. As I pointed out in ref 2, the actual vacancy excitations are undoubtedly severely renormalized from the Hartree-Fock theory and have a finite mass which is NOT that introduced later. We here are interested in the ground state.

The localized wave functions ϕ(r) are by no means orthogonal to each other, and there are no points at which the density vanishes. If the solid forms, undoubtedly the overlaps between them will be small, but never zero. On the other hand, it is important to realize that the potential well arises as a consequence of the fact that the particle cannot interact with itself: it is there because each site contains exactly one particle which is repelled by its neighbors but not by its own potential. The hole-particle gap occurs when there is exactly one particle per site. The Bose solid is what I called a "true" solid, equivalent to an insulator in the electronic analog, in that there is a downward cusp in the energy as a function of occupancy of the sites at exactly one—or an integer number—per site.

We can express this requirement in a rigorous way as a requirement on the particle current *as distinct from motion of the site locations*. That is, we define a current $\mathcal{J}$ of bosons (helium atoms) *relative to the lattice of sites*, and the condition on this current is that

$$\nabla \cdot \mathcal{J} \equiv 0 \qquad [3]$$

Equation [3] is distinct from the property of rigidity which is the standard definition of a solid, and expresses the concept of "true solid", which I was trying clumsily to elucidate in reference 2, as distinct from a three-dimensional density wave. It is in general true of most classical solids as far as the atoms are concerned, but only for insulating solids when these are considered as density waves of electrons. It expresses the fact that in this sense our model is equivalent to the Bose Hubbard model[3] in the "insulating" state.

But equation [3] does not imply that $\mathcal{J}=0$, necessarily. Let us consider our Bose solid of equation [1]. Unlike the Bose Hubbard model, where the sites are treated as independent boson operators, the nonorthogonal boson operators c* are not independent and do not commute with each other: they can interfere constructively or destructively. When the lattice as a whole moves, part of the resulting current will be carried in the overlap region, and the magnitude of that current will depend on the relative phases of the boson fields. The insulating nature of the state means that absolute phase has no meaning, but as far as I can see relative phase can and does have meaning. Of course, the current has to be instantly replenished from other sites in order to satisfy [3], but this is not impossible for superfluid flow. The contacts (superleaks) between the atom sites provide a stiffness energy depending on the phase difference between them, completely analogous to a Josephson energy. The velocity associated with this flow is surely not the standard superfluid expression $v \neq (\hbar/m)\nabla\phi$, which depends on local Galilean invariance, but something considerably smaller.

We define a local relative phase $\phi$ coarse-grained on the scale of the atoms, and the above arguments tell us that there will be some coarse-grained expression for the current:

$$\mathcal{J} = \text{const} \times \nabla\phi \qquad [4]$$

which need not express Galilean invariance because the lattice serves as a preferred reference frame: the constant (which I don't attempt to calculate) represents a small fraction of the total mass, and is proportional to the overlap between neighboring φ's. This may be very roughly estimated in terms of the exchange energy in solid He3, which is of the order of the overlap squared times an interaction energy. We can arbitrarily define the constant in [4] in terms of a heavy "effective mass" M by

$$v = \frac{\hbar}{M}\nabla\phi \qquad [5]$$

It is, however, essential that when the lattice of sites and its superfluid fraction (which is what [4] is) are set in uniform motion, the whole mass move with them. Thus when there is a constant gradient of the phase, [5] and the "normal" fraction J of the current, that which is carried along with the sites, must add up to the total mass of helium—the normal fraction is less than unity.

Because of [4] the supercurrent has a second constraint:

$$\nabla \times \mathcal{J} = 0 \qquad [6]$$

Of course, if [3] and [6] are satisfied everywhere, then the current can only be 0, or at most a constant; but if we allow for a line defect where we can make Ψ=0, that is we break a one-dimensional manifold of our superleaks, we can satisfy them with a vortex flow with v∝1/r.

Now if we try a rotation experiment, the rigid rotation of the site lattice does not obey $\nabla \times v = 0$, so that in the absence of a vortex singularity the superfluid fraction cannot participate in the rotation at absolute zero. As we raise the rotation velocity, at low temperatures vortices will be drawn into the sample in order to mimic rigid rotation a la Onsager-Feynman. There will be a critical angular velocity, analogous to $H_{c1}$ of a type II superconductor, when the rotation energy is first equal to the cost of vortices. I estimate that because the vorticity unit of the superflow is small($\propto 1/M$), this is quite small, of the order of one quantum of ordinary vorticity, and comparable to the observed threshold. (The smallness of 1/M cancels the large logarithm in the vortex energy.)

Beyond the threshold the superfluid fraction will drop logarithmically. As the temperature rises, thermally activated vortices will gradually appear and allow the solid to equilibrate in the rigidly rotating state; the transition region will be dissipative like the vortex liquid of ordinary superconductors. The genuinely rigid, solid phase is a liquid of free vortices, in which forces will be accommodated by vortex motion or, equivalently, free phase slippage, rather than by superleak currents. Isotopic impurities will nucleate vortices and thus will tend to destroy the superfluidity and restore rigidity.

It seems that this scenario is compatible with all the observations so far, although of course it badly needs a more quantitative approach. One interesting additional experiment has been suggested to me[4]: The supercurrent should not couple very well to longitudinal phonons, since for these $\nabla \cdot J \approx 0$, while it should be very visible for transverse ones. In this it differs very much from liquid He, where the superflow is of true particles and couples well to phonons.

The main point here is that the definition [3] of the criterion for a true solid is not incompatible with the existence of a superfluid fraction for Bose solids. It is significant that the familiar rigid solid turns out to be actually in yet another way an emergent phenomenon.

I would like to acknowledge extensive discussions with M H W Chan and with W F Brinkman, without which this paper would not have been written.

[1] E Kim and M. H W Chan, Nature 427, 225, (2004); Science 305, 1941, (2005)
[2] P W Anderson, "Basic Notions of Condensed Matter Physics", Ch 4 pp143 et seq, Benjamin, Menlo Park CA, 1984
[3] M F W Fisher, et al , Phys Rev B40, 546 (1989)
[4] W F Brinkman, private communication